\documentclass[12pt]{article}
\usepackage{}
\usepackage{amsfonts}
\usepackage{bbm}
\usepackage{bm}
\usepackage{mathrsfs}
\usepackage{color}
\usepackage{geometry}             
\usepackage{graphicx}
\usepackage{amssymb}
\usepackage{amsmath}
\usepackage{cite}
\usepackage{comment}
\usepackage{tikz}
\usetikzlibrary{matrix}

\usepackage[hyperindex=true,
          pdfstartview=FitH,
          bookmarksnumbered=true,
          bookmarksopen=true,
          citecolor=blue,
          linkcolor=blue,
          colorlinks=true,
          urlcolor=rossoCP3,
          pdfborder=001,
          unicode]{hyperref}

\definecolor{rossoCP3}{cmyk}{0,.88,.77,.40}

\DeclareMathOperator{\arcosh}{arcosh}

\newcommand{\rd}[1]{\mathrm{d}#1}

\newcommand{\re}{\mathrm{e}}

\newcommand{\rR}{\mathrm{R}}

\newcommand{\g}{\mathfrak{g}}

\newcommand{\Li}{\mathrm{Li}}

\newcommand{\pfrac}[2]{\left(\frac{\partial #1}{\partial #2}\right)}

\usepackage{xcolor}  
\definecolor{mygreen}{RGB}{44,85,17}
\definecolor{myblue}{RGB}{34,31,217}
\definecolor{mybrown}{RGB}{194,164,113}
\definecolor{myred}{RGB}{255,66,56}
\definecolor{mypurple}{RGB}{200,36,176}

\newcommand{\blue}[1]{\textcolor{blue}{#1}}
\renewcommand{\blue}[1]{\textcolor{black}{#1}}

\allowdisplaybreaks

\begin{document}

\title{Gravito-thermal transports, Onsager reciprocal relation
and gravitational Wiedemann-Franz law}
\author{Xin Hao${}^a$, Song Liu${}^b$,
and Liu Zhao\thanks{Corresponding author.}\hspace{.5em}${}^b$
\vspace{1pt}\\
\small ${}^a$School of Physics, Hebei Normal University, Shijiazhuang 050024, China\\
\small ${}^b$School of Physics, Nankai University, Tianjin 300071, China\\
\small {email}: \href{mailto:xhao@hebtu.edu.cn}{xhao@hebtu.edu.cn},\\
\small \href{mailto:2120190124@mail.nankai.edu.cn}{2120190124@mail.nankai.edu.cn}
\small and \href{mailto:lzhao@nankai.edu.cn}{lzhao@nankai.edu.cn}}

\date{}

\maketitle
\begin{abstract}
Using the near-detailed-balance distribution function obtained in
our recent work, we present a set of covariant gravito-thermal transport
equations \blue{(i.e. the flow of various charges as linear response to 
thermodynamical forces)} for neutral relativistic gases in a generic stationary spacetime.
All relevant tensorial transport coefficients are worked out and are presented
using some particular integration functions in $(\alpha,\zeta)$, where
$\alpha = -\beta\mu$ and $\zeta =\beta m$ is the relativistic coldness,
with $\beta$ being the inverse temperature and $\mu$ being the chemical potential.
It is shown that the Onsager reciprocal relation holds in the gravito-thermal
transport phenomena, and that the heat conductivity and the gravito-conductivity tensors
are proportional to each other, with the coefficient of proportionality given by
the product of the so-called Lorenz number with the temperature,
thus proving a gravitational variant of the Wiedemann-Franz law.
It is remarkable that, for strongly degenerate Fermi gases, the Lorenz number
takes a universal constant value $L=\pi^2/3$, which extends the Wiedemann-Franz law
into the Wiedemann-Franz-Lorenz law.
\end{abstract}

\section{Introduction}

A salient feature of thermodynamics is that of universality. Different systems
can have similar qualitative macroscopic behaviors
irrespective of the microscopic structures. This feature is especially
manifest for systems under equilibrium. Beyond that, in the case when
slight departures from global equilibrium appear, it is common for all fluid
systems to share similar phenomenological transport laws and heavy
emphasis was placed on the general symmetry principle that
restricts the kinetic coefficients, namely the Onsager reciprocal relation
\cite{PhysRev.37.405,PhysRev.38.2265}.

Historically, a great deal of efforts have been devoted to the unveiling of
universality in irreversible processes, and the general framework develops
over long periods of time until various empirical relations can be reformulated using the
methods from statistical mechanics \cite{RevModPhys.17.343,APERTET2016225}.
However, in this pattern, attention has rarely been
paid to the transport phenomena controlled fully by relativistic gravitational
field. The reason for such a situation stems from the fact that, on one hand,
in laboratory condensed matter experiments, gravity is
too weak to play a key role, on the other hand, the relativistic
statistical mechanics is far from being established.

Over the past few years, with the abundant data accumulated by the LIGO/Virgo
gravitational wave detectors \cite{LIGOScientific:2016aoc} and the Event Horizon Telescope
\cite{Akiyama_2019},
physics in strong gravity becomes ready to merit broader attentions.
This new era of fundamental physics and astronomy calls for an extension of
thermodynamics and statistical mechanics to strong gravity regime.
On the theoretical side, a statistical mechanical explanation of black hole
thermodynamics is expected to contribute insight into quantum gravity
\cite{Padmanabhan:2009vy,Almheiri:2020cfm}.
For the observational studies, the brightest sources in the universe
are powered black holes, where the temperature of the near-horizon
environment reaches $10^9$ K and much of the gravitational potential
energy is converted into heat, which is in turn carried away by
radiation \cite{Narayan:2023ivq}.
To this end, it becomes increasingly important to provide additional understanding
about the thermodynamic behaviors of the relativistic gases that are
flowing around black holes. The aim of the present work is to study the transport
phenomena of relativistic gases subjecting to strong gravity.

The earliest investigation about gravito-thermal effects was carried out
by Tolman and Ehrenfest \cite{PhysRev.35.904,PhysRev.36.1791},
who noticed a remarkable feature of gravity.
Building on the idea that heat --- as a form of energy --- should also
gravitate, they argued that
the temperature gradient could be compensated for by
gravitational field to restore equilibrium, and vice versa.
This mechanism is known as the Tolman-Ehrenfest (TE) effect,
and since the original outset, both its validity and generality have been
extensively studied \cite{PhysRev.76.427.2,RevModPhys.21.531,Rovelli:2010mv,
Lima:2019brf,PhysRevD.105.L081501,Majhi:2023stp,Xia:2023idh}.
The most famous application of the TE effect is in the
Luttinger theory of thermal transport coefficients \cite{PhysRev.135.A1505},
where gravity comes into play as a counter-term for the temperature gradient,
which enables the calculation of thermal transport coefficients
within the framework of linear response theory. This is the beginning for
gravity to appear in condensed-matter physics, and the seminal work of Luttinger
has now been generalized to the quantum level, where various new transport effects
arise due to anomalies \cite{Chernodub:2021nff}.
In all these cases, only weak gravitational field is needed. Please be reminded that
there is a parallel effect which applies to the gradient of the chemical potential,
known as the Klein effect. Although these gravito-thermal effects are still
too weak to be observed directly, the simplest way to unify the description for
various transport phenomena is to acknowledge their existence in principle.

Another well-known approach to transport phenomena is kinetic theory
\cite{Jutter,PhysRev.122.1342,israel1963}
which has been successfully generalized in a form that is manifestly covariant,
and increasingly applied to the study of quark-gluon plasma
\cite{PhysRevLett.51.351,GABBANA20201,Hidaka:2022dmn}, cosmology and
black hole accretion \cite{Bazow:2015dha,Husdal:2016pfd,Pordeus-da-Silva:2019bak,
Sasankan:2019oee, Rioseco:2016jwc,Mach:2021zqe,Acuna-Cardenas:2021nkj}.
Recently in \cite{Liu:2022wpu}, taking advantage of
the relativistic Boltzmann equation and an observer dependent collision model, we
obtained a covariant transport equation for the particle number flow by collecting the
TE effect and Klein effect in the generalized gradients of the temperature and chemical
potential, and obtained in analytic form the corresponding tensorial transport
coefficients, including the gravito-conductivity tensor as a particular example.

The present paper is a continuation of the previous study \cite{Liu:2022wpu}.
The main purpose is three-folded. Firstly, as further applications of the
near-equilibrium solution found in \cite{Liu:2022wpu}, we take
a step forward to construct the transport equations for the internal energy
and entropy flows and calculate the corresponding transport tensors in covariant form.
Due to the fundamental local thermodynamic relation which still holds in systems
obeying the local equilibrium assumption, the entropy flow is not independent of the
particle number and internal energy flows. Secondly, assembling the
transport coefficients for the particle number and internal energy
flows together, we verify that the
celebrated Onsager reciprocal relation still holds for the kinetic coefficients
associated with the gravito-thermal transport phenomena. Lastly,
by introducing the heat flow as the internal energy flow
in the absence of net particle number flow, we obtain the heat conductivity
tensor, which is then verified to be proportional to the gravito-conductivity,
with product of the temperature and the so-called Lorenz number playing
the role of proportionality factor. This proves a gravitational variant of the
Wiedemann-Franz law. The Lorenz number $L$ is described analytically by some
expression involving several integration functions,
and, interestingly, for strongly degenerate Fermi gases, $L$
takes a universal constant value $\pi^2/3$ in both the non-relativistic
and ultra-relativistic limits.

\section{Thermodynamic forces in gravitational field}

This section is a reminder for the basics of conventional non-equilibrium thermodynamics
and relativistic kinetic theory. We shall also introduce notations and quantities
which are necessary for describing non-equilibrium thermodynamic processes in
curved spacetimes.

Non-equilibrium systems contain the flows of particle number and various other
quantities (``charges'') driven by thermodynamic forces. For near equilibrium systems, the
flows can be viewed as linear responses to the thermodynamic forces, and the
response factors are known as transport coefficients. For example, in the absence
of particle number flow, Fourier's law relates the heat flow to temperature gradient;
while in the absence of heat flow, Fick's law describes the relation between
particle diffusion and concentration gradient. Non-equilibrium thermodynamics
provides us with a basis for understanding these phenomenological laws.

To deal with non-equilibrium systems, the first difficulty arises in defining the entropy.
This problem can be solved when the characteristic time of macroscopic evolution
is much larger than the timescale for microscopic process but otherwise
much smaller than the relaxation time for the system to become equilibrated.
In such cases, the local equilibrium assumption could be adopted, which assumes that,
although the global fundamental relation of equilibrium thermodynamics is necessarily
broken, the local variant in terms of densities should still hold,
\begin{align}
\mathrm{d} s =  \frac{1}{T}\mathrm{d}\epsilon - \frac{\mu}{T}\mathrm{d}n,
\label{ds}
\end{align}
where $s$,  $\epsilon$ and $n$ respectively represent the local densities
of the entropy, internal energy and particle number. The local fundamental relation
as presented above is usually referred to as the entropic representation.
In this representation, the intensive quantities are known as the Massieu parameters,
i.e.
\begin{align*}
\alpha = \frac{\partial s}{\partial n}= -\frac{\mu}{T},
\qquad \beta =\frac{\partial s}{\partial \epsilon} = \frac{1}{T},
\end{align*}
and their gradients are defined as the entropic thermodynamic forces
\begin{align}
\vec{\mathcal F}^{(\alpha)} = \nabla \alpha, \qquad
\vec{\mathcal F}^{(\beta)} = \nabla \beta.
\label{TFNR}
\end{align}
Later, in the relativistic context, the parameter $\beta$ is often replaced by
the dimensionless parameter $\zeta=\beta m$, which is called the relativistic coldness.

In the presence of non-vanishing thermodynamic forces,
eq.\eqref{ds} suggests the following relation between the entropy
flow $\vec j_s$ with the internal energy flow $\vec j_\epsilon$ and
the particle number flow $\vec j_n$,
\begin{align}
\vec j_s = \alpha \vec j_n + \beta \vec j_\epsilon.
\label{jsNR}
\end{align}
Consequently, the entropy production rate will be
expressed in terms of flows and the thermodynamic forces,
\begin{align*}
\mathcal G_s = \vec{\mathcal F}^{(\alpha)} \cdot \vec j_n
+ \vec{\mathcal F}^{(\beta)} \cdot \vec j_\epsilon.
\end{align*}
To close the above equations, one needs supplementary transport equations
relating the flows to thermodynamic forces. A common approach is to expand the flows
in powers of the thermodynamic forces, and for near equilibrium systems
it is sufficient to restrict the discussion to the linear order. As a result, one has
\begin{align*}
\bm j^a =
\left( \begin{matrix}
		j^a_n \\
		j^a_\epsilon\\
\end{matrix} \right)
=
\left( \begin{matrix}
	L^{\,ab}_{(\alpha\alpha)}&	L^{\,ab}_{(\alpha\beta)}\\
	L^{\,ab}_{(\beta\alpha)}&	L^{\,ab}_{(\beta\beta)}\\
\end{matrix} \right)
\left( \begin{matrix}
		\mathcal F_b^{(\alpha)} \\
		\mathcal F_b^{(\beta)} \\
\end{matrix} \right)
= \bm L^{ab} \bm{\mathcal F}_b.
\end{align*}
The components of matrix $\bm L^{ab}$ are called kinetic coefficients and
they capture the characteristics of non-equilibrium systems in the linear response
regime. These kinetic coefficients can be obtained by experiment or
by means of non-equilibrium statistical mechanics. Either way, there is an
symmetry principle for $\bm L^{ab}$. For charged system, and in the presence of
external magnetic field $\bm{B}$, the Onsager relation is given by
\begin{align}
\bm L^{ab}\left[\bm{B}\right]
= \left(\bm L^{\mathrm{T}}\right)^{ba}\left[-\bm{B}\right],
\end{align}
which is of fundamental importance in non-equilibrium thermodynamics.

The above brief picture for non-equilibrium thermodynamics did not take the microscopic
description (i.e. statistical physics description) into account, and has nothing to do
with gravity. In order to present a statistical description for a neutral relativistic
fluid subjecting strong gravity, we employ the relativistic kinetic theory, in which
the one particle distribution function (1PDF) $f$ plays a key role.
In the following, it is suffices to consider only the primary variables,
namely the particle number current $N^\mu$, the energy momentum tensor $T^{\mu\nu}$
and the entropy current $S^\mu$, for the relativistic fluid. These objects
are determined by 1PDF $f$ and the metric
tensor $g_{\mu\nu}$ via
\begin{align}
&N^\mu= \int \bm\varpi p^\mu f, \qquad
T^{\mu\nu}= \int\bm\varpi p^\mu p^\nu f,  \nonumber\\
&S^\mu = - \int \bm\varpi p^\mu
\left[f \log f - \varsigma^{-1}
\left(1 + \varsigma f \right)
\log \left(1 + \varsigma f\right)\right], \label{NTS}
\end{align}
where $\bm\varpi = \g \frac{\sqrt{g}}{p_0} (\rd p)^d$ is invariant
volume element on the momentum space with $\g$ being the intrinsic degeneracy factor,
and $\varsigma=0,+1,-1$ corresponds respectively to the Maxwell-Boltzmann,
Bose-Einstein and Fermi-Dirac statistics.
Since the integrations are performed only over the momentum space,
$T^{\mu\nu}$, $N^\nu$ and $S^\mu$ are local quantities in spacetime.
Please be reminded that, throughout this paper, we will work in units
$k_{\rm B}=c=\hbar =1$ and the convention on the signature of the metric
is the mostly positive one.

Strictly speaking, the dynamics of $g_{\mu\nu}$ and $f$ are governed by a set of
coupled equations including Einstein equation and Boltzmann equation, but
the present work considers only probe systems in a prescribed spacetime background,
thus $f$ is to be regarded as the solution to the relativistic Boltzmann
equation
\[
\pounds_H f(x,p)=\mathcal C(x,p),
\]
where $\pounds_H$ represents the Liouville vector field on the tangent bundle
over the fixed $(d+1)$-dimensional curved spacetime background, and $\mathcal C(x,p)$
represents the collision term.

In fact, an exact treatment of the relativistic Boltzmann equation alone is still
perplexing because it is hard to determine the collision term in closed form.
In the cases when the collision contribution completely disappears, the Boltzmann equation
degenerates into the Liouville equation, and there is a unique solution $\bar f$
under the assumptions that the elementary processes are time-reversal invariant and
the entropy production rate vanishes. $\bar f$ is known as the detailed balance
distribution and it takes the form
\begin{align}
\bar f= \frac{1}{\re^{\bar \alpha- \bar{\mathcal B}_\mu p^\mu} - \varsigma},
\label{fbar}
\end{align}
where $\bar\alpha$ and $\bar{\mathcal B}^\mu$ are constrained by
\begin{align}
\partial_\mu \bar \alpha = 0, \qquad \qquad \nabla_{(\mu} \mathcal{\bar B}_{\nu)} = 0.
\label{DBcondition}
\end{align}
By convention, all ``barred'' variables refer to quantities describing
systems in detailed balance.
Notice that the equation obeyed by $\bar{\mathcal B}^\mu$ is a Killing equation,
and in order to have a non-relativistic limit, the expression $- \bar{\mathcal B}_\mu p^\mu$
needs to be proportional to the energy of the particle, and thus requires
$\bar{\mathcal B}_\mu$ to be timelike. Therefore, in order for the detailed balance
distribution \eqref{DBcondition} to exist, $\bar{\mathcal B}^\mu$ is required to be a
timelike Killing vector field, which in turn implies that the spacetime background
must be at least stationary.

The primary variables $N^\mu$, $T^{\mu\nu}$ and $S^\mu$ have phenomenological
interpretations. As such, their component forms must be observer dependent.
For systems in stationary spacetimes, the common choice for the observers is
that of the stationary class, whose proper velocity field $Z^\mu$ is proportional to
the timelike Killing vector field $\bar{\mathcal B}^\mu$,
\begin{align*}
\mathcal{\bar B}^\mu = \bar \beta Z^\mu,\qquad Z^\mu Z_\mu =-1.
\end{align*}
In terms of such observers, the primary variables for the relativistic fluid
under detailed balance are evaluated to be
\begin{equation}
\begin{split}
&\bar{N}^\mu = \g m^d \mathcal A_{d-1} \bar J_{d-1,1}\, Z^\mu, \\
&\bar T^{\mu\nu} = \g m^{d+1} \mathcal A_{d-1}
\left(\bar J_{d-1,2} \,Z^\mu Z^\nu +\frac{1}{d} \bar J_{d+1,0}
\, \Delta^{\mu\nu} \right), \\
&\bar S^\mu =  \g m^d \mathcal A_{d-1}
\left(\bar\alpha \bar J_{d-1,1} + \bar\zeta \bar J_{d-1,2}
+\frac{\bar\zeta}{d} \bar J_{d+1,0}\right) Z^\mu ,
\end{split}    \label{NTSbar}
\end{equation}
where $\mathcal A_{d-1}$ is the area of the $(d-1)$-dimensional unit sphere,
$m$ is the mass of constituent particles,
$\Delta^{\mu\nu} =g^{\mu\nu}+ Z^\mu Z^\nu$ is the projection tensor,
and $\bar J_{n,m}$ represents the following function in
$(\bar\alpha,\bar\zeta)$, with $\bar\zeta = \bar\beta m$,
\begin{align}
&\bar  J_{n,m} = J_{n,m}(\bar \alpha, \bar \zeta,\varsigma),\nonumber\\
\qquad
&J_{n,m}(\alpha,\zeta,\varsigma) \equiv \int_0^\infty\frac{\sinh^n \vartheta
\cosh^m\vartheta}{\re^{\alpha+ \zeta \cosh \vartheta} - \varsigma}
\rd \vartheta.
\label{Jnm}
\end{align}
The form of eq.\eqref{NTSbar} implies that the stationary observers are
automatically comoving for systems in detailed balance.

From equations \eqref{DBcondition} and \eqref{NTSbar} we can conclude that:

\begin{description}
\item[i)] In the eyes of stationary observer, the system under detailed
balance is completely characterized by the scalar densities
$\bar n$, $\bar\epsilon$ and $\bar s$ and the pressure $\bar P$,
which are all functions in $\bar \alpha$ and $\bar \beta$,
\begin{align}
\bar n &\equiv - Z_\mu N^\mu = \g m^d \mathcal A_{d-1}  \bar J_{d-1,1},
\quad\qquad
\bar \epsilon \equiv Z_\mu Z_\nu T^{\mu\nu}
= \g m^{d+1} \mathcal A_{d-1}\bar J_{d-1,2}, \\
\bar P &\equiv \frac{1}{d}T^{\rho\sigma}\Delta_\rho{}^\mu \Delta_{\sigma\mu}
= \g m^{d+1} \frac{\mathcal A_{d-1}}{d} \bar J_{d+1,0},
\quad\,\,
\bar s \equiv -Z_\mu S^\mu
=  \bar \alpha \bar n + \bar \beta \bar \epsilon
+ \bar \beta \bar P.
\end{align}

\item[ii)] In differential form, we have the relations
\begin{align}
\rd \bar s = \bar \alpha \, \rd \bar n + \bar \beta\,\rd \bar \epsilon,
\qquad  \rd \bar P
= \bar s \, \rd (\bar \beta^{-1}) - \bar n \, \rd (\bar \alpha \bar \beta^{-1}).
\end{align}
Therefore, $\bar \alpha$ and $\bar \beta$ acquire the interpretation as Massieu parameters
under detailed balance,
\[
\bar\alpha=-\bar\mu/\bar T,\qquad \quad  \bar\beta=1/\bar T.
\]
The parameter $\bar\zeta = \bar\beta m$ then is understood to be the relativistic
coldness in detailed balance.

\blue{
\item[iii)] The system under detiled balance does not involve expansion or shear effect
\begin{align}
  \nabla_{\nu} Z^{\nu} = 0, \qquad
  \Delta^{\mu \rho} \Delta^{\nu \sigma} \nabla_{(\rho } Z_{ \sigma)} -
  \frac{1}{d} \nabla_{\rho} Z^{\rho} {\Delta}^{\mu \nu} = 0.
  \label{expansion&shear}
\end{align}
}

\item[iv)] There is no macroscopic transports in systems under detailed balance.
However, the gradients of the temperature and chemical potential can still be nonvanishing,
\begin{align}
\nabla_\nu \bar T = - \bar T Z^\sigma\nabla_\sigma Z_\nu,
\qquad
\nabla_\nu \bar \mu = - \bar \mu  Z^\sigma\nabla_\sigma Z_\nu.
\label{TEK}
\end{align}
If the right hand sides do not vanish, these gradients can be understood as manifestations
of the TE and Klein effects.
\end{description}

In view of the last property, it is proposed in \cite{Liu:2022wpu} that
the gradients of the temperature and chemical potential should be generalized as
\begin{align}
\mathcal D_\nu T \equiv \nabla_\nu T
+ T Z^\sigma\nabla_\sigma Z_\nu,
\qquad
\mathcal D_\nu  \mu \equiv \nabla_\nu  \mu
+ \mu Z^\sigma\nabla_\sigma Z_\nu.
\end{align}
For any function $\Psi$ in $(T,\mu)$, we also introduce the corresponding
generalized gradient
\begin{align}
\mathcal D_\nu \Psi(T,\mu)
=\pfrac{\Psi}{T}\mathcal D_\nu T+\pfrac{\Psi}{\mu}\mathcal D_\nu \mu,
\label{defchain}
\end{align}
and define
\begin{align}
\dot \Psi = Z^\nu\mathcal D_\nu \Psi.
\label{defdot}
\end{align}
Thus the generalized gradients of the Massieu parameters $\alpha$ and $\beta$ are given by
\begin{align}
\mathcal{D}_\nu\alpha = \mathcal{D}_\nu\left(-\frac{\mu}{T}\right)
= \nabla_\nu \alpha,
\quad
\mathcal{D}_\nu\beta = \mathcal{D}_\nu\left(\frac{1}{T}\right)
= \nabla_\nu \beta - \beta Z^\sigma\nabla_\sigma Z_\nu,
\label{DA&DB}
\end{align}
which, according to equations \eqref{TEK}, are both identically
zero under detailed balance.

When $\mathcal{D}_\nu\alpha$ and $\mathcal{D}_\nu\beta$ are nonvanishing,
they can be further decomposed into two parts, i.e. the parallel and
orthogonal parts with respect to $Z_\nu$,
\begin{align*}
\mathcal{D}_\nu\alpha = \dot \alpha \, Z_\nu + \mathcal F^{(\alpha)}_\nu,
\qquad
\mathcal{D}_\nu\beta = \dot \beta \, Z_\nu + \mathcal F^{(\beta)}_\nu,
\end{align*}
where the thermodynamic forces $\mathcal F^{(\alpha)}_\mu$ and
$\mathcal F^{(\beta)}_\mu$ are defined as the spacelike vectors
\begin{align}
\mathcal F^{(\alpha)\mu}= \Delta^{\mu\nu} \mathcal D_\nu \alpha,
\qquad
\mathcal F^{(\beta)\mu} = \Delta^{\mu\nu} \mathcal D_\nu \beta
=\frac{1}{m}\Delta^{\mu\nu} \mathcal D_\nu \zeta.
\label{TFR}
\end{align}
We shall see later that such decompositions are crucial for separating the
spacelike transport flows from the spacetime currents. Notice that, in our
terminology, the spacelike vector parts of the spacetime vector fields
such as $N^\mu, S^\mu$ are referred to as {\it flows}, while $N^\mu, S^\mu$ themselves
are {\it currents}.

\section{Perturbations and transport equations}

To drive the system out of equilibrium, let us turn on the generalized gradients
of Massieu parameters, i.e. set $\mathcal D_\nu \alpha \neq 0$, $\mathcal D_\nu \beta \neq 0$.
Assuming that the departure from detailed balance is not too severe,
it is reasonable to approximate the 1PDF $f$,
in the zeroth order, by an expression which is similar in form to the
detailed balance distribution \eqref{fbar},
\begin{align}
f^{(0)} =  \frac{\g}{\re^{ \alpha- \mathcal B_\mu p^\mu} - \varsigma}.
\label{led}
\end{align}
This approximation is referred to as the local equilibrium distribution.
The difference between the local equilibrium distribution and the detailed balance
distribution lies in that, there is no constraint for $\alpha$ any more and
$\mathcal{B}^\mu$ is no longer required to be a Killing vector field.
In the present work, we confine ourselves to the case when $\mathcal{B}^\mu$
differs from $\bar{\mathcal{B}}^\mu$ only in magnitude but not in direction,
therefore, $\mathcal{B}^\mu$ is still proportional to the proper velocity $Z^\mu$
of the stationary observer,
\[
\mathcal B^\mu\equiv \beta Z^\mu.
\]

To the zeroth order, the local properties of the system are very similar to the case
when the system is under detailed balance, and the constitutive relations are
\begin{align}
\left[N^{(0)}\right]^\mu = n^{(0)} Z^\mu, \quad
\left[T^{(0)}\right]^{\mu\nu}
= \epsilon^{(0)} Z^\mu Z^\nu + P^{(0)} \Delta_{\mu\nu}, \quad
\left[S^{(0)}\right]^\mu = s^{(0)} Z^\mu,
\label{ZO}
\end{align}
where
\begin{align}
~~~~~~~~~
n^{(0)} &= \g m^{d} \mathcal A_{d-1} J_{d-1,1},
&& \epsilon^{(0)} = \g m^{d+1} \mathcal A_{d-1} J_{d-1,2}, \nonumber\\
P^{(0)} &= \g m^{d+1}\frac{\mathcal A_{d-1}}{d} J_{d+1,0},
&& s^{(0)} = \alpha n^{(0)} + \beta \epsilon^{(0)}
+ \beta P^{(0)}.
~~~~~~~~~\label{nePs-0rd}
\end{align}
The zeroth order densities given above are all functions in $(\alpha, \beta)$.

Since the thermodynamic forces did not appear in eq.\eqref{ZO}, there is no
transport flows in the zeroth order. In order to study the transport
phenomena, it is essential to find the correction $\delta f$ to the local equilibrium
distribution to the first order in the thermodynamic forces.
Using an observer dependent collision model proposed
in \cite{Liu:2022wpu}, $\delta f$ is found to be proportional to thermodynamic forces,
\begin{align}
\delta f = - \frac{\tau}{\beta\varepsilon}\frac{\partial f^{(0)}}{\partial \varepsilon}
p^\mu\left(\delta_\mu{}^\nu + \frac{\tau}{\varepsilon}
\Delta^{\alpha}{}_\mu \nabla_{[\alpha}p_{\beta]}\,\Delta^{\beta\nu} \right)
\left(\mathcal D_\nu \alpha + \varepsilon \,\mathcal D_\nu \beta \right),
\label{deltaf}
\end{align}
where $\tau$ is the relaxation time, and $\varepsilon = -Z_\mu p^\mu$ is the
energy of the particle, as measured by the observer $Z^\mu$.
This in turn implies that the corrections to the particle number
current $N^\mu$, the energy momentum tensor $T^{\mu\nu}$ and the entropy current
$S^\mu$ are all proportional to thermodynamic forces given in eq.\eqref{TFR}.

Before we proceed with the detailed calculations, let us introduce
some notations needed in the following sections.

While studying non-inertial fluids or systems in curved spacetimes,
the formulae can be simplified in appropriate coordinates.
Sometimes the coordinate vector field is not hypersurface orthogonal.
In such cases, the rotation of the corresponding coordinate line can be
described by an antisymmetric tensor. For example, in the basis
with $Z^\mu = \left(e_{\hat 0}\right)^\mu$, we can introduce
\begin{align}
B_{\mu\nu} = \nabla_{[\alpha} Z_{\beta]}
\Delta^\alpha{}_\mu \Delta^\beta{}_\nu,
\label{B2}
\end{align}
to describe the rotation of $\left(e_{\hat 0} \right)^\mu$, which, in the post-Newtonian
limit,  corresponds to the gravitomagnetic (GM) field. For the
rotation of the other coordinate vector fields, all the relevant parts in
the present work will be encoded in the following tensor
\begin{align}
\mathfrak{B}_{\sigma\mu\nu}
=\mathfrak{B}_{\sigma[\mu\nu]}
= (e_{\hat a})_\sigma \, \omega^{\hat a}{}_{[\alpha\beta]}
\Delta^\alpha{}_\mu \Delta^\beta{}_\nu
\label{B3}
\end{align}
where
\begin{align*}
\omega^{\hat \alpha}{}_{\hat \beta \mu} =
-\left[\partial_\mu (e^{\hat \alpha})_\nu- \Gamma^{\rho}{}_{\mu\nu}
(e^{\hat \alpha})_\rho \right](e_{\hat \beta})^\nu,
\end{align*}
is the spin connection. Please keep in mind that
$B_{\mu\nu}$ and $\mathfrak B_{\mu\nu\sigma}$ are given in terms of the
first derivatives of the metric tensor and basis vectors. We also introduce the notation
\[
\mathfrak B_{\mu}= g^{\rho\sigma} \mathfrak B_{\rho\sigma\mu}.
\]

The first order corrections to $N^\mu, T^{\mu\nu}$ and $S^\mu$ are given as follows,
and we refer to Appendix A for details,
\begin{align}
\delta N^\mu
= & ~ \tau \g m^{d}  \frac{\mathcal A_{d-1}}{d}
\bigg[d\mathcal J_{d-1,1}{Z^\mu Z^\nu}
+ \tau \mathcal J_{d+1,-1} {Z^\mu} \mathfrak{B}^\nu
+ \mathcal J_{d+1,-1} \left(\Delta^{\mu\nu} + \tau B^{\mu\nu}
\right) \bigg] \mathcal D_\nu \alpha \nonumber\\
&+ \tau \g m^{d} \frac{\mathcal A_{d-1}}{d}
\bigg[d \mathcal J_{d-1,2}{Z^\mu Z^\nu}
+ \tau \mathcal J_{d+1,0} {Z^\mu} \mathfrak{B}^\nu
+ \mathcal J_{d+1,0} \left(\Delta^{\mu\nu} + \tau B^{\mu\nu}
\right) \bigg] \mathcal D_\nu \zeta, \\
\delta T^{\mu\nu}
= & ~ \tau \g m^{d+1} \frac{\mathcal A_{d-1}}{d}
\bigg\{ d \mathcal J_{d-1,2}{Z^\mu Z^\nu Z^\sigma}
+ \tau \mathcal J_{d+1,0} {Z^\mu Z^\nu} \mathfrak{B}^\sigma \nonumber\\
&\qquad + \mathcal J_{d+1,0} \left[\Delta^{\mu\nu}{Z^\sigma}
+{Z^\mu} \left(\Delta^{\nu\sigma} + \tau B^{\nu\sigma} \right)
+{Z^\nu} \left(\Delta^{\mu\sigma} + \tau B^{\mu\sigma} \right)\right] \nonumber\\
&\qquad + \frac{\tau}{d+2} \mathcal J_{d+3,-2}
\left(\Delta^{\mu\nu} \mathfrak B^\sigma
+ \mathfrak B^{\mu\nu\sigma} + \mathfrak B^{\nu\mu\sigma} \right) \bigg\} \mathcal D_\sigma \alpha \nonumber\\
&+ \tau \g m^{d+1} \frac{\mathcal A_{d-1}}{d}
\bigg\{d\mathcal J_{d-1,3}{Z^\mu Z^\nu Z^\sigma}
+ \tau \mathcal J_{d+1,1} {Z^\mu Z^\nu} \mathfrak{B}^\sigma \nonumber\\
&\qquad + \mathcal J_{d+1,1}\left[\Delta^{\mu\nu}{Z^\sigma}
+{Z^\mu} \left(\Delta^{\nu\sigma} + \tau B^{\nu\sigma} \right)
+{Z^\nu} \left(\Delta^{\mu\sigma} + \tau B^{\mu\sigma} \right)\right] \nonumber\\
&\qquad + \frac{\tau}{d+2} \mathcal J_{d+3,-1}
\left(\Delta^{\mu\nu} \mathfrak B^\sigma
+ \mathfrak B^{\mu\nu\sigma} + \mathfrak B^{\nu\mu\sigma} \right)
\bigg\}\mathcal D_\sigma \zeta, \\
\delta S^\mu
= & ~\tau \g m^{d} \frac{\mathcal A_{d-1}}{d}
\bigg[d \mathcal K_{d-1,1}{Z^\mu Z^\nu}
+ \tau \mathcal K_{d+1,-1} {Z^\mu} \mathfrak{B}^\nu
+ \mathcal K_{d+1,-1} \left(\Delta^{\mu\nu} + \tau B^{\mu\nu}
\right) \bigg] \mathcal D_\nu \alpha  \nonumber \\
& + \tau \g m^{d}  \frac{\mathcal A_{d-1}}{d}
\bigg[d \mathcal K_{d-1,2}{Z^\mu Z^\nu}
+ \tau \mathcal K_{d+1,0} {Z^\mu} \mathfrak{B}^\nu
+ \mathcal K_{d+1,0} \left(\Delta^{\mu\nu} + \tau B^{\mu\nu}\right) \bigg]
\mathcal D_\nu \zeta,
\label{decomp1}
\end{align}
where
$\mathcal J_{n,m}$, $\mathcal K_{n,m}$ are functions in $(\alpha, \zeta)$,
which are defined below,
\begin{align}
\mathcal J_{n,m}(\alpha,\zeta)
&= -\frac{\partial}{\partial \alpha} J_{n,m}(\alpha,\zeta,\varsigma)
=-\frac{\partial}{\partial \zeta} J_{n,m-1}(\alpha,\zeta,\varsigma)\nonumber\\
&=- \frac{\partial}{\partial \zeta}  \int_0^\infty\frac{\sinh^n \vartheta
\cosh^{m-1}\vartheta}{\re^{ \alpha+\zeta \cosh \vartheta} - \varsigma}
\rd \vartheta, \\
\mathcal K_{n,m}(\alpha,\zeta)
&= - \left(\alpha \frac{\partial}{\partial \alpha}
+\beta \frac{\partial}{\partial \beta}\right) J_{n,m}(\alpha,\zeta,\varsigma)
\nonumber\\
&=\alpha \mathcal{J}_{n,m}(\alpha,\zeta) + \zeta \mathcal J_{n,m+1}(\alpha,\zeta).
\label{JK}
\end{align}

We can further decompose $\delta N^\mu, \delta T^{\mu\nu}$ and $\delta S^\mu$ as
\begin{align}
&\delta N^\mu = \delta n \, Z^\mu + j_n^\mu,  \qquad
\delta S^\mu = \delta s \, Z^\mu + j_s^\mu,\nonumber\\
& \delta T^{\mu\nu} = \delta \epsilon\, Z^\mu Z^\nu
+ j_\epsilon^\mu Z^\nu
+ j_\epsilon^\nu Z^\mu + \delta P \Delta^{\mu\nu}
+ \Pi^{\mu\nu},
\label{decomp2}
\end{align}
where $j^\mu_n, j^\mu_\epsilon, j^\mu_s$ and $\Pi^{\mu\nu}$ are all orthogonal to $Z^\mu$,
\[
Z_\mu j^\mu_n=Z_\mu j^\mu_\epsilon=Z_\mu j^\mu_s=0,\qquad
Z_\mu \Pi^{\mu\nu}=\Pi^{\mu\nu} Z_\nu =0.
\]
Then the scalar, vector and tensor parts of the above decomposition can be read off
by comparing \eqref{decomp1} with \eqref{decomp2}.

\begin{description}
\item[i)] Scalar parts: $\delta n$, $\delta \epsilon$, $\delta P$
and $\delta s$ are respectively higher-order corrections to the particle
number density, energy density, pressure and entropy density,
\begin{align}
\delta n &= -\tau \dot{n}^{(0)}
+\tau^2 {\g}{m^d} \frac{\mathcal A_{d-1}}{d} \mathfrak{B}^\nu
\left(\mathcal J_{d+1,-1} \mathcal D_\nu \alpha
+\mathcal J_{d+1,0}\,\mathcal D_\nu \zeta \right), \\
\delta \epsilon &= -\tau \dot{\epsilon}^{(0)}
+\tau^2 {\g}m^{d+1} \frac{\mathcal A_{d-1}}{d} \mathfrak{B}^\nu
\left(\mathcal J_{d+1,0} \mathcal D_\nu \alpha
+\mathcal J_{d+1,1}\,\mathcal D_\nu \zeta \right), \\
\delta P &= -\tau \dot{P}^{(0)}
+\tau^2 {\g}m^{d+1} \frac{\mathcal A_{d-1}}{d^2} \mathfrak{B}^\nu
\left(\mathcal J_{d+3,-2} \mathcal D_\nu \alpha
+\mathcal J_{d+3,-1}\,\mathcal D_\nu \zeta \right), \\
\delta s &= -\tau \left(\alpha \dot{n}^{(0)}
+ \beta \dot{\epsilon}^{(0)}  \right)
+\tau^2 {\g}m^{d} \frac{\mathcal A_{d-1}}{d} \mathfrak{B}^\nu
\left(\mathcal K_{d+1,-1} \mathcal D_\nu \alpha
+\mathcal K_{d+1,0}\,\mathcal D_\nu \zeta \right),
\end{align}
where all dotted objects are defined as in eq.\eqref{defdot}.
It is easy to show that the identity
\begin{align}
\delta s = \alpha \delta n + \beta \delta \epsilon
\end{align}
holds which is in accordance with the fundamental relation \eqref{ds}.

\item[ii)] Vector parts: $j_n^{\mu}$, $j_\epsilon^\mu$ and $j_s^\mu$
are respectively the particle number flow, internal energy flow and
the entropy flow,
\begin{align}
j_n^\mu &= \tau {\g}m^{d} \frac{\mathcal A_{d-1}}{d}
\left(\Delta^{\mu\nu} + \tau B^{\mu\nu}\right)
\left(\mathcal J_{d+1,-1} \mathcal D_\nu \alpha
+\mathcal J_{d+1,0}\,\mathcal D_\nu \zeta \right), \label{jn} \\
j_\epsilon^\mu &=
\tau {\g}m^{d+1} \frac{\mathcal A_{d-1}}{d}
\left(\Delta^{\mu\nu} + \tau B^{\mu\nu}\right)
\left(\mathcal J_{d+1,0}\, \mathcal D_\nu \alpha
+ \mathcal J_{d+1,1}\, \mathcal D_\nu \zeta \right), \label{je} \\
j_s^\mu &= \tau {\g}m^{d} \frac{\mathcal A_{d-1}}{d}
\left(\Delta^{\mu\nu} + \tau B^{\mu\nu}\right)
\left(\mathcal K_{d+1,-1} \mathcal{D}_\nu \alpha
+\mathcal K_{d+1,0}\,\mathcal D_\nu \zeta \right). \label{js}
\end{align}
Eqs.\eqref{jn}-\eqref{js} represent the sought-for covariant transport equations,
which give a complete description for the vector transport phenomena driven by the
thermodynamic forces $\mathcal D_\mu \alpha$ and $\mathcal D_\mu \zeta$.
Invoking the relation \eqref{JK}, the entropy flow $j_s^\mu$ can be related to
$j_n^\mu$ and $j_\epsilon^\mu$ via
\begin{align}
j_s^\mu = \alpha j_n^\mu + \beta j_\epsilon^\mu,
\end{align}
which agrees with eq.\eqref{jsNR}.

\item[iii)] The tensor part is due to the non-inertial effect, where
$\Pi^{\mu\nu}$ is traceless and symmetric,
\begin{align}
\Pi^{\mu\nu} = \tau^2 {\g}m^{d+1} \frac{\mathcal A_{d-1}}{d(d+2)}
\left(\mathfrak B^{\mu\nu\sigma} + \mathfrak B^{\nu\mu\sigma}
-\frac{2}{d}\Delta^{\mu\nu}\mathfrak{B}^\sigma \right)
\left(\mathcal{J}_{d+3,-2} \mathcal{D}_\sigma \alpha
+\mathcal{J}_{d+3,-1} \mathcal{D}_\sigma \zeta\right).
\end{align}
\blue{It is important to emphasize that in our setting only the Massieu parameters $\alpha$ and $\beta$,
or equivalently the temperature $T$ and chemical potential $\mu$, are perturbed.
The direction of $\mathcal B^\mu$, or in other words the proper velocity $Z^\mu$
of the comoving observer, remains unchanged. Therefore, according to 
eq.~\eqref{expansion&shear}, no shear and viscous effects could arise 
under such perturbations. However, due to the non-inertial effects, 
such as the spin of observer,
an apparent deviatoric stress still could be observed.
By highlighting that only the Massieu parameters are perturbed in our specific setting,
we will, in the current study, focus on the the thermodynamic force associated with these parameters.}

\end{description}

Finally, let us concentrate on eqs.\eqref{jn} and \eqref{je}. These two equations can be
rearranged into a more compact form
\begin{align}
\bm j^\mu =
\left( \begin{matrix}
		j^\mu_n \\
		j^\mu_\epsilon\\
\end{matrix} \right)
=
\left( \begin{matrix}
	L^{\,\mu\nu}_{(\alpha\alpha)}&	L^{\,\mu\nu}_{(\alpha\beta)}\\
	L^{\,\mu\nu}_{(\beta\alpha)}&	L^{\,\mu\nu}_{(\beta\beta)}\\
\end{matrix} \right)
\left( \begin{matrix}
		\mathcal F_\nu^{(\alpha)} \\
		\mathcal F_\nu^{(\beta)} \\
\end{matrix} \right)
= \bm L^{\mu\nu} \bm{\mathcal F}_\nu.
\end{align}
In doing so, all kinetic coefficients are encoded in $\bm L^{\mu\nu}$,
\begin{align}
\bm L^{\mu\nu} = \tau \g m^d \frac{\mathcal A_{d-1}}{d}
\left(\Delta^{\mu\nu} + \tau B^{\mu\nu}\right)
\left( \begin{matrix}
	\mathcal J_{d+1,-1}&	m \mathcal J_{d+1,0}\\
	m \mathcal J_{d+1,0}&	m^2 \mathcal J_{d+1,1}\\
\end{matrix} \right). \label{kincoef}
\end{align}
It can be easily seen that the Onsager reciprocal relation is fulfilled by
the kinetic coefficients associated with the gravito-thermal transport phenomena,
\begin{align}
\bm L^{\mu\nu}\left[B\right]
= \left(\bm L^{\mathrm{T}}\right)^{\nu\mu}\left[-B\right],
\label{Onsgrav}
\end{align}
wherein $B$ is a shorthand for $B^{\mu\nu}$.

\blue{
To better demonstrate the influence of gravitational fields on transport phenomena,
and illustrate more clearly the gravito-thermal effect, let us 
consider a probe system in the Schwarzschild spacetime with line element
\[
\rd s^2=-\left(1-\frac{2GM}{r}\right)\rd t^2+\left(1-\frac{2GM}{r}\right)^{-1}\rd r^2
+r^2\rd \Omega^2.
\]
The set of stationary observers is chosen to be
\[
Z^\mu = \left(1-\frac{2GM}{r}\right)^{-1/2} \left(\partial_t\right)^\mu.
\]
Correspondingly, we have the following generalized gradients,
\begin{align}
\mathcal D_\nu \alpha = \partial_\nu \alpha, \quad
\mathcal D_\mu \beta = \partial_\mu \beta - \beta Z^\sigma\nabla_\sigma Z_\nu
= \partial_\mu \beta - \beta\frac{GM}{r^2} \left(\partial_r\right)_\mu,
\end{align}
and also
\begin{align}
B_{\mu\nu} = \nabla_{[\alpha} Z_{\beta]}
\Delta^\alpha{}_\mu \Delta^\beta{}_\nu = 0.
\end{align}
In this case, the non-zero comments of the particle number flow and internal
energy flow are
\begin{align}
j_n^{\,a} &= \tau {\g}m^{d} \frac{\mathcal A_{d-1}}{d}
\left(\mathcal J_{d+1,-1} \partial^a \alpha
+\mathcal J_{d+1,0}\,\partial^a \beta
- \beta\mathcal J_{d+1,0}\frac{GM}{r^2} \left(\partial_r\right)^a \right), \label{jne.g.1} \\
j_\epsilon^\mu &=
\tau {\g}m^{d+1} \frac{\mathcal A_{d-1}}{d}\left(\mathcal J_{d+1,0} \partial^a \alpha
+\mathcal J_{d+1,1}\,\partial^a \beta
- \beta\mathcal J_{d+1,1}\frac{GM}{r^2} \left(\partial_r\right)^a \right), \label{jee.g.1}
\end{align}
where $a = r, \varphi, \theta$. 
In eqs.~\eqref{jne.g.1} and \eqref{jee.g.1}, the first two terms are
the usual contribution from the spatial gradient of Massieu parameters, while the last terms
indicate the nontrivial gravito-thermal coupling. Let us remark that the contribution 
of gravitational force to the transport fluxes has already been reported in Ref.\cite{Kremer}.
}

Before closing this section, let us mention that the gravito-thermal transports
phenomena have already been studied in previous works using relativistic kinetic theory
with some different collision models. However, the role of observer was not exploited
upon, and the kinetic coefficients were presented either in a component by component
manner or in terms of some finitely cut off polynomials. We refer to
the books \cite{degroot1980relativistic,cercignani2002relativistic} for details.
Our construction has the merits that manifest general covariance is
kept throughout, the role of observer is properly encoded, and that the tensorial
kinetic coefficients are presented in fully analytic form in terms of several
integration functions. It is precisely these extra merits that enables the direct
verification of the Onsager reciprocal relation \eqref{Onsgrav}.

\section{Wiedemann-Franz law and Lorenz number}

Having calculated the kinetic coefficients in the previous section,
various phenomenological transport coefficients can be derived immediately.
As an example, we shall deduce the gravitational conductivity,
heat conductivity and discuss their relations.

For this purpose, it is customary to rewrite the particle number flow and
the internal energy flow in terms of the generalized gradients of
the chemical potential, $\mathcal D_\nu \mu$, and that of the
temperature, $\mathcal D_\nu T$. We have
\begin{align}
j_n^\mu &= - {\tau} \g m^{d-1} \frac{\mathcal A_{d-1}}{d}
\left(\Delta^{\mu\nu} + \tau B^{\mu\nu}\right) \zeta
\left[\mathcal J_{d+1,-1} \mathcal D_\nu \mu
+ \left(\alpha\mathcal J_{d+1,-1}
+\zeta\mathcal J_{d+1,0}\right) \mathcal D_\nu T \right],  \\
j_\epsilon^\mu &=-{\tau} \g m^{d}  \frac{\mathcal A_{d-1}}{d}
\left(\Delta^{\mu\nu} + \tau B^{\mu\nu}\right) \zeta
\left[\mathcal J_{d+1,0} \mathcal D_\nu \mu
+ \left(\alpha\mathcal J_{d+1,0}
+\zeta\mathcal J_{d+1,1}\right) \mathcal D_\nu T \right].  \label{JeT}
\end{align}
The gravito-conductivity $\sigma^{\mu\nu}$ can be read off from $j^\mu_n$ by
turning off the generalized temperature gradient, i.e. setting $\mathcal D_\nu T = 0$,
which gives
\begin{align}
j^\mu_n &= \sigma^{\mu\nu} \mathcal D_\nu \mu, \qquad
\sigma^{\mu\nu} =
-{\tau} \g m^{d-1}  \frac{\mathcal A_{d-1}}{d}
\left(\Delta^{\mu\nu} + \tau B^{\mu\nu}\right)
\zeta \mathcal J_{d+1,-1}.
\label{grc}
\end{align}
The heat flow is defined to be the internal energy flow in the absence
net particle number flow. In such cases, there is a balance
between $\mathcal D_\nu \mu$ and $\mathcal D_\nu T$
and the ratio is defined as the gravitational Seebeck coefficient,
\begin{align}
\mathcal D_\nu \mu
= -\left(\alpha+\zeta \frac{\mathcal J_{d+1,0}}{\mathcal J_{d+1,-1}}\right)
\mathcal D_\nu T.   \label{SC}
\end{align}
Inserting eq.\eqref{SC} into eq.\eqref{JeT}, the elimination of $\mathcal D_\nu \mu$
gives the transport equation for heat flow
\begin{align}
q^\mu &=
\kappa^{\mu\nu} \mathcal D_\nu T, \quad
\kappa^{\mu\nu} = - {\tau} \g m^{d}\frac{\mathcal A_{d-1}}{d}
\left(\Delta^{\mu\nu} + \tau B^{\mu\nu}\right)
\zeta^2 \frac{\mathcal J_{d+1,-1}\mathcal J_{d+1,1}
-\mathcal J_{d+1,0}^2}{\mathcal J_{d+1,-1}},
\label{htc}
\end{align}
where $\kappa^{\mu\nu}$ is the heat conductivity tensor.

Inspecting the expressions for the gravito-conductivity tensor $\sigma^{\mu\nu}$
\eqref{grc} and the heat conductivity tensor $\kappa^{\mu\nu}$ \eqref{htc},
one can easily find the following relation,
\begin{align}
\kappa^{\mu\nu}= L T \sigma^{\mu\nu}, \label{WF}
\end{align}
where
\begin{align}
L= \zeta^2 \frac{\mathcal J_{d+1,-1}\mathcal J_{d+1,1}
-\mathcal J_{d+1,0}^2}{\mathcal J_{d+1,-1}^2}.
\label{Lorenz}
\end{align}
Eq.\eqref{WF} is the gravitational analogue of the well-known Wiedemann-Franz law,
whose original version describes a proportionality relationship between
the heat conductivity and the electric conductivity --- both are scalar transport
coefficients --- in metals \cite{Wilson:1953,Price}. Now the gravitational variant of the
Wiedemann-Franz law is established as a relationship between two tensorial
transport coefficients. The factor $L$ given in eq.\eqref{Lorenz} is henceforth
referred to as the Lorenz number, as in the original version of the Wiedemann-Franz law.
In the present case, it is evident that the Lorenz number
is in generical {\it not} a constant, but rather a function in $(\alpha,\zeta)$.

There are a couple of important limiting cases in which the expression for the Lorenz number
can be simplified drastically. Especially,
in the non-relativistic $(\zeta \gg 1)$ and the ultra-relativistic $(\zeta \ll 1)$
limits, we find a unified formula
\begin{align}
L = \ell(\ell+1)\frac{\Li_{\ell+1} \left(\varsigma z\right)}
{\Li_{\ell -1}\left(\varsigma z\right)}
-\ell^2\frac{\Li_\ell\left(\varsigma z\right)^2}
{\Li_{\ell-1}\left(\varsigma z\right)^2},
\label{LL}
\end{align}
where $\Li_\nu(x)$ is the polylogarithm function
\[
\Li_\nu(x) = \sum_{k=1}^{\infty} \frac{x^k}{k^\nu},
\]
$\ell$ depends on the spacetime dimension
\begin{align}
\ell & = d/2+1, ~~~~~~\, (\zeta\gg1)   \label{ellnr} \\
\ell & = d,     ~~~~~~~~~~~~~~~ (\zeta\ll1)  \label{ellur}
\end{align}
and $z\equiv \re^{-\alpha_*}$ with $\alpha_* = -\beta \mu_* =
-\beta(\mu - m) = \alpha + \zeta$ is the fugacity which
characterizes the degree of degeneracy.

Assuming that eq.\eqref{LL} holds, and if we further
consider the non-degenerate limit, i.e. $\alpha_* \to \infty$,
there will be no difference between Fermi and Bose gases.
The Lorenz number in such cases can be further simplified and we have
\begin{align*}
z \to 0
\quad \Rightarrow \quad
\Li_\nu \left(\varsigma z\right) \approx \varsigma z
\quad \Rightarrow \quad
L \to \ell.
\end{align*}

On the other hand, strongly degenerate Fermi and Bose gases behave quite differently,
and the difference is also reflected in the Lorenz number.

The chemical potential of a Bose gas
is non-positive, $\mu_* \leqslant 0$, thus the strongly degenerate Bose gas
is characterized by $\alpha_* \to 0$. In this case, we have
\begin{align*}
z \to 1
~ \Rightarrow ~
L \to \ell(\ell+1)\frac{\zeta_\rR\left(\ell+1\right)}{\zeta_\rR\left(\ell-1\right)}
-\ell^2\frac{\zeta_\rR\left(\ell\right)^2}{\zeta_\rR\left(\ell-1\right)^2},
\end{align*}
where $\ell$ takes values as given in eq.\eqref{ellnr} or eq.\eqref{ellur}, and
$\zeta_{\rm R}(x)$ denotes the Riemann $\zeta$ function, which should not
be confused with the relativistic coldness $\zeta$.
In all cases mentioned above, the Lorenz number is dependent on $\ell$ and hence
on the spacetime dimension, which leads to different results in
the ultra-relativistic and non-relativistic limits.

The strongly degenerate Fermi gas, however, has completely different
behavior. For such systems, we can insert the strongly degenerate condition
$\alpha_* \to -\infty$ into the original expression \eqref{Lorenz} for the Lorenz number
and make simplifications right from there. The result is surprisingly simple and universal,
\begin{align}
z \to \infty
~ \Rightarrow ~
L \to \frac{\pi^2}{3}.
\label{LFermi}
\end{align}
This result depends neither on the dimension and geometry of the underlying
spacetime, nor on the relativistic coldness $\zeta$.
If the SI units were adopted, the above universal limiting value should read
$\displaystyle L=\frac{\pi^2}{3}(k_{\rm B})^2$. Please be reminded that,
the gravito-conductivity $\sigma^{\mu\nu}$ is associated with the particle number flow.
If it were associated with the mass flow, there would be an additional factor
depending on the mass $m$ of the particle, and the above universal value for
strongly degenerate Fermi gas becomes $\displaystyle L=\frac{\pi^2}{3}
\left(\frac{k_{\rm B}}{m}\right)^2$.
In contrast, in the original Wiedemann-Franz law describing
the relationship between heat and electric conductivities for charged non-relativistic
Fermi gases, the Lorenz number takes the value $\displaystyle
L= \frac{\pi^2}{3}\left(\frac{k_{\rm B}}{e}\right)^2$ for strongly degenerate Fermi gas
\cite{Wilson:1953}, where $e$ represents the electron charge. The surprising
similarity between the Lorenz numbers for strongly degenerate Fermi gases
subjecting to electric field and gravity respectively is quite remarkable.
The Lorenz numbers \eqref{LL} in various limiting cases are shown
graphically in Figure 1. The horizontal axis in the right
plots is intentionally reversed to make sharper contrast between the
non-relativistic and ultra-relativistic limits for non-degenerate systems.

\begin{figure}[h]
	\begin{center}
		\includegraphics[height=.35\textheight]{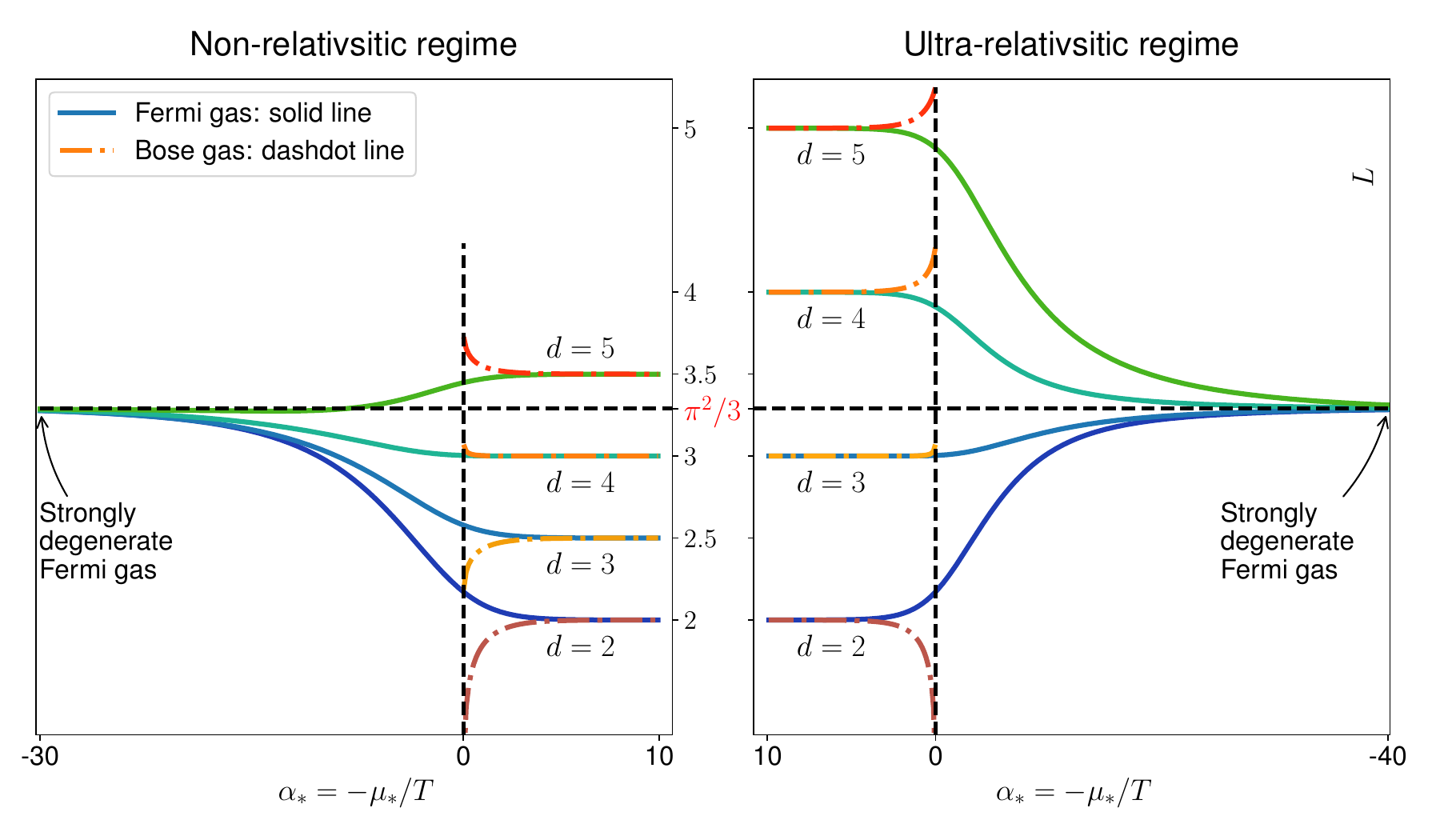}	
		
		\caption{The Lorenz number}
	\end{center} \label{PLL}
\end{figure}

\section{Concluding remarks}

Guided by the viewpoint that observer is the key to bridge the gap between
thermodynamics and relativity\cite{Hao:2021ifw},
the prior work presented a new collision model for
relativistic Boltzmann equation and obtained a near equilibrium distribution function
which is best used for studying transport phenomena in stationary spacetimes.
The advantage of this distribution is illustrated by particle number transport
where the effect of relativistic gravitational field is quite apparent.
In this study, further examples are given to test the effectiveness of the distribution.
The covariant transport equations for all the primary variables in
the framework of relativistic kinetic theory are derived, surprisingly,
the best interpretation of the results is rooted in Onsager theory for
non-equilibrium processes. More concretely, both particle number flow and the
internal energy flow are proportional to the thermodynamic forces, which,
are identified as the generalized gradients of Massieu parameters. In this way,
the kinetic coefficients are verified to obey the Onsager reciprocal relation
which originally arose in non-relativistic statistical physics context.

The present work is built on top of relativistic kinetic theory and a specific,
observer dependent collision model. The results are succinct and highly consistent with
the conventional non-equilibrium thermodynamics. In addition, we proved the
gravitational variant of the Wiedemann-Franz law and calculated the Lorenz
number for neutral systems in gravitational field. Most notably, the Lorenz
number for strongly degenerate Fermi gas takes a universal value $\pi^2/3$,
which depends neither on the dimension and geometry of the spacetime, nor on the
relativistic coldness. This result extends the Wiedemann-Franz law into
the Wiedemann-Franz-Lorenz law \cite{Price}.

It needs to be stressed that the present work considers only the transport equations for
neutral relativistic systems subjecting to gravity. For astrophysical applications,
the transport equations for charged relativistic systems subjecting to
both gravitational and electromagnetic fields will be more appealing. Moreover,
kinetic theory is not the only approach to non-equilibrium physics, some of the
assumptions and models are well justified, but some approximations are applied
for convenience and these must be checked against complementary statistical methods
such as the stochastic processes, see, e.g.
\cite{debbasch2004diffusion,dunkel2009relativistic,herrmann2010diffusion,
haba2010relativistic,Cai:2023cqr}. Nevertheless, we hope our preliminary treatment
can be a stepping-stone in such areas.

Let us close this paper by adding the remark that, although we have chosen
the stationary observer throughout this paper, the same construction should also
work for other choices of observers. If other observers were considered, then
the thermodynamic parameters for the system in detailed balance should get transformed
in a way as discussed in \cite{Hao:2021ifw}. Accordingly, the transport tensors
should also acquire a transformation. Such transformations are {\it not}
coordinate transformations but rather arise as the consequences of the choice of observers.
This leaves us with an interesting question
as to whether the form of the Onsager reciprocal relation and the Wiedemann-Franz law
are observer independent. We hope to come back to this question in a later study.

\appendix

\section{Corrections to the primary variables \label{Appendix A}}

We are free to choose any orthonormal basis $\{(e_{\hat \alpha})^\mu\}$,
satisfying $\eta_{\hat\alpha\hat\beta} =g_{\mu\nu}(e_{\hat \alpha})^\mu
(e_{\hat \beta})^\nu$, 
in the tangent space at a given spacetime event, and
in order to simplify the computation, it is convenient to fix the timelike leg such that
$Z^\mu = (e_{\hat 0})^\mu$. Then the projection tensor is simply given by
\begin{align*}
\Delta^{\mu\nu} = g^{\mu\nu} + Z^\mu Z^\nu = \delta^{\hat a \hat b}
(e_{\hat a})^\mu (e_{\hat b})^\nu.
\end{align*}
In this basis, the component of the momentum,
$p^{\hat \alpha} = p^\mu (e^{\hat \alpha})_\mu$,
can be parameterized by a real rapidity parameter $\vartheta$ and a spacelike unit vector
$n^{\hat a} \in S^{d-1}$ as
\begin{align*}
p^{\hat \alpha} &= m\left(\cosh \vartheta, ~ \sinh \vartheta \, n^{\hat a}\right).
\end{align*}
The momentum space volume element then reduces to
\begin{align*}
\bm\varpi &= \g (m \sinh\vartheta)^{d-1}
\rd \vartheta \, \rd \Omega_{d-1},
\end{align*}
with $\rd \Omega_{d-1}$ being the volume element of the $(d-1)$-dimensional
unit sphere $S^{d-1}$. Therefore, the momentum space integral can be split
into two parts, i.e. the integral over $\vartheta\in(0,\infty)$ and that
over $S^{d-1}$. For the first part, we have introduced the function $J_{n,m}$
in eq.\eqref{Jnm}, and, for the second part, we note that
\begin{align*}
&\int n^{\hat a}\rd\Omega_{d-1}= 0,  \qquad \qquad
\int n^{\hat a}n^{\hat b}\rd\Omega_{d-1}=\frac{1}{d}\mathcal{A}_{d-1}\delta^{\hat a\hat b},    \\
&\int n^{\hat a}n^{\hat b}n^{\hat c} \rd\Omega_{d-1}=0,  \qquad
\int n^{\hat a}n^{\hat b}n^{\hat c}n^{\hat d} \rd\Omega_{d-1}
=\frac{1}{d}\mathcal{A}_{d-1}\delta^{\hat a\hat b\hat c\hat d},
\end{align*}
where $ \mathcal A_{d-1}= \displaystyle\int \rd\Omega_{d-1}$
is the area of a $(d-1)$-dimensional unit sphere,
$\delta^{\hat a\hat b}$ is the Kronecker symbol and
\begin{align*}
\delta^{\hat a \hat b \hat c \hat d}
\equiv \frac{1}{d+2}\left(\delta^{\hat a \hat b}\delta^{\hat c \hat d}
+ \delta^{\hat a \hat c}\delta^{\hat b \hat d}
+ \delta^{\hat a \hat d}\delta^{\hat b \hat d}\right),
\qquad \delta^{\hat a \hat b \hat c \hat d}\,\delta_{\hat c \hat d}
= \delta^{\hat a \hat b}.
\end{align*}

The purpose of this appendix is to derive the expressions for
$\delta N^\mu$, $\delta T^{\mu\nu}$
and $\delta S^\mu$ in terms of $(\alpha, \zeta, Z^\mu)$, where, by definition,
\begin{align}
N^\mu &= \left[N^{(0)}\right]^\mu + \delta N^\mu
=  \int \bm\varpi p^\mu f^{(0)} + \int \bm\varpi p^\mu \delta f , \\
T^{\mu\nu} &= \left[T^{(0)}\right]^{\mu\nu} + \delta T^{\mu\nu}
= \int \bm\varpi p^\mu p^\nu f^{(0)}
+ \int \bm\varpi p^\mu p^\nu \delta f,
\end{align}
and
\begin{align}
S^\mu = \left[S^{(0)}\right]^\mu + \delta S^\mu,
\end{align}
with
\begin{align}
&S^\mu = -\int \bm\varpi p^\mu
\left[f \log f - \varsigma^{-1}
\left(1 + \varsigma f \right)
\log \left(1 + \varsigma f\right)\right],\\
\left[S^{(0)}\right]^\mu &= - \int \bm\varpi p^\mu
\left[f^{(0)} \log f^{(0)} - \varsigma^{-1}
\left(1 + \varsigma f^{(0)} \right)
\log \left(1 + \varsigma f^{(0)}\right)\right],
\end{align}
and $f=f^{(0)}+\delta f$.  Thinking of $\delta f$ as a small correction to $f^{(0)}$,
one can easily get
\begin{align}
\delta S^\mu &= -\int \bm\varpi p^\mu\, \delta f \log\frac{f^{(0)}}{1+\varsigma f^{(0)}}
= \alpha\,\delta N^\mu + \mathcal B_\nu \,\delta T^{\mu\nu}.
\end{align}

In order to outline the derivation of eq.\eqref{decomp1}, we need to recall that
$\delta f$ is a combination of terms of orders $\mathcal O(\tau)$ and $\mathcal O(\tau^2)$.
Therefore, the correction terms to be evaluated should also be calculated up to the order
$\mathcal O(\tau^2)$. As intermediate results, let us list the following
equations and identities,
\begin{align*}
\mathcal D_\nu f^{(0)}
=\frac{1}{\beta}\frac{\partial f^{(0)}}{\partial \varepsilon}
\left(\mathcal D_\nu \alpha + \varepsilon \,\mathcal D_\nu \beta \right),
\end{align*}
\begin{align*}
\int\bm\varpi \, p^\mu f^{(0)}
=& \mathcal \g m^{d} \mathcal A_{d-1} J_{d-1,1} \, Z^\mu , \\
\int\bm\varpi \, p^\mu p^\nu f^{(0)}
=&\g m^{d+1} \mathcal A_{d-1}
\left(J_{d-1,2} \,Z^\mu Z^\nu +\frac{1}{d} J_{d+1,0}
\, \Delta^{\mu\nu} \right), \\
\int\bm\varpi \, p^\mu p^\nu p^\sigma f^{(0)}
=&\g m^{d+2} \mathcal A_{d-1} \left(J_{d-1,3} Z^\mu Z^\nu Z^\sigma
+ \frac{3}{d} J_{d+1,1} \Delta^{(\mu\nu} Z^{\sigma)}\right), \\
\Delta^\alpha{}_\nu \int\bm\varpi \, p^\mu p^\nu n^{\hat a} f^{(0)}
=& \g m^{d+1} \frac{\mathcal A_{d-1}}{d} J_{d,1}Z^\mu (e^{\hat a})^\alpha, \\
\Delta^\alpha{}_\sigma \int\bm\varpi \, p^\mu p^\nu p^\sigma n^{\hat a} f^{(0)}
=& \g m^{d+2} \frac{\mathcal A_{d-1}}{d}  \left[
J_{d,2}\,Z^\mu Z^\nu (e^{\hat a})^\alpha
+ \frac{3}{d+2} J_{d+2,0}\, \Delta^{(\mu\nu}(e^{|\hat a|})^{\alpha)} \right].
\end{align*}
The rest calculations are straightforward. At the order $\mathcal{O}(\tau)$, we have
\begin{align}
\delta N^\mu\left[\mathcal{O}(\tau)\right]
=& - \frac{\tau}{m}\mathcal D_\nu
\int \bm \varpi \,\frac{p^\mu p^\nu}{\cosh\vartheta} f^{(0)} \nonumber\\
=& - \tau \g m^{d}\mathcal A_{d-1}
\left(Z^\mu Z^\nu \mathcal D_\nu J_{d-1,1}
+ \frac{1}{d} \Delta^{\mu\nu} \mathcal D_\nu J_{d+1,-1} \right), \\
\delta T^{\mu\nu} \left[\mathcal{O}(\tau)\right]
=& - \frac{\tau}{m}\mathcal D_\sigma
\int \bm \varpi \, \frac{p^\mu p^\nu p^\sigma}{\cosh\vartheta} f^{(0)}\nonumber \\
=& - \tau \g m^{d+1} \mathcal A_{d-1}
\left(Z^\mu Z^\nu Z^\sigma \mathcal D_\sigma J_{d-1,2}
+ \frac{3}{d} \Delta^{(\mu\nu} Z^{\sigma)} \mathcal D_\sigma J_{d+1,0} \right),
\end{align}
where
\begin{align}
\mathcal{D}_\sigma J_{m,n}
= \frac{\partial J_{m,n}}{\partial \alpha} \mathcal D_\nu \alpha
+ \frac{\partial J_{m,n}}{\partial \zeta} \mathcal D_\nu \zeta
= -\mathcal{J}_{m,n} \mathcal{D}_\sigma \alpha
-\mathcal{J}_{m,n+1} \mathcal{D}_\sigma \zeta.
\end{align}
Next, at the order $\mathcal{O}(\tau^2)$, we have
\begin{align}
\delta N^\mu\left[\mathcal{O}(\tau^2 )\right]
=& -\frac{\tau^2}{m} \Delta^\alpha{}_{\sigma}
\nabla_{[\alpha} Z_{\beta]} \Delta^{\beta\nu} \mathcal D_\nu
\int \bm \varpi \,\frac{p^\mu p^\sigma}{\cosh\vartheta} f^{(0)} \nonumber\\
&-\frac{\tau^2}{m} \Delta^\alpha{}_{\sigma}
\nabla_{[\alpha} (e_{|\hat a|})_{\beta]} \Delta^{\beta\nu} \mathcal D_\nu
\int \bm \varpi \,\frac{p^\mu p^\sigma n^{\hat a} \sinh\vartheta}{\cosh^2\vartheta} f^{(0)} 
\nonumber\\
=& - \tau^2  \g m^{d} \frac{\mathcal A_{d-1}}{d}
\left(B^{\mu\nu} + Z^\mu \mathfrak B^\nu \right)\mathcal D_\nu J_{d+1,-1},
\end{align}
\begin{align}
\delta T^{\mu\nu}\left[\mathcal{O}(\tau^2 )\right]
=& -\frac{\tau^2}{m} \Delta^\alpha{}_\rho
\nabla_{[\alpha} Z_{\beta]} \Delta^{\beta\sigma} \mathcal D_\sigma
\int \bm \varpi \,\frac{p^\mu p^\nu p^\rho}{\cosh\vartheta} f^{(0)} \nonumber\\
&-\frac{\tau^2}{m} \Delta^\alpha{}_\rho
\nabla_{[\alpha} (e_{|\hat a|})_{\beta]} \Delta^{\beta\sigma} \mathcal D_\sigma
\int \bm \varpi \,\frac{p^\mu p^\nu p^\rho n^{\hat a} \sinh\vartheta}{\cosh^2\vartheta} f^{(0)} \nonumber\\
=& - \tau^2 \g m^{d+1} \frac{\mathcal A_{d-1}}{d}
\left(Z^{(\mu}B^{\nu)\sigma} + Z^\mu Z^\nu \mathfrak B^\sigma \right)
\mathcal D_\sigma J_{d+1,0} \nonumber\\
& - \tau^2 \g m^{d+1} \frac{\mathcal A_{d-1}}{d(d+2)}
\left(\Delta^{\mu\nu} \mathfrak B^\sigma
+ \mathfrak B^{\mu\nu\sigma} + \mathfrak B^{\nu\mu\sigma} \right)
\mathcal D_\sigma J_{d+3,-2},
\end{align}
where $B_{\mu\nu}$ and $\mathfrak B^{\sigma}{}_{\mu\nu}$ are defined as in
eqs.\eqref{B2} and \eqref{B3}. Finally, summing up contributions of
orders $\mathcal O(\tau)$ and $\mathcal O(\tau^2)$ together and
rearranging terms, we obtain the results listed in eq.\eqref{decomp1}.

\section{Lorenz number in the cases $\zeta\gg1$ and $\zeta\ll1$
\label{Appendix B}}

The aim of this section is to simplify Lorenz number \eqref{Lorenz}
in the limits $\zeta\gg1$ and $\zeta\ll1$.  Therefore,
we take $n=d+1$ and $m= -1, 0 ,1$.

In terms of the variable $u = \zeta(\cosh\vartheta -1)$, we have
\begin{align*}
\cosh\vartheta = \dfrac{u}{\zeta}+1, \quad
\sinh\vartheta = \sqrt{\frac{u^2}{\zeta^2}+2\frac{u}{\zeta}}, \quad
\rd \vartheta = \frac{\rd u}{\zeta \sinh\vartheta},
\end{align*}
and the integral $J_{n,m}$ becomes
\begin{align}
J_{n,m} = \frac{1}{\zeta} \int_0^\infty
\frac{\left(\frac{u^2}{\zeta^2}+2\frac{u}{\zeta}\right)^{{n-1}\over{2}}
\left(\frac{u}{\zeta}+1\right)^m}{\re^{\alpha_*+u}-\varsigma} \rd u.
\end{align}

In the non-relativistic limit $\zeta \gg 1$,
$J_{n,m}$ is approximately
\begin{align}
J_{n,m} &\approx \frac{1}{2}\left(\frac{2}{\zeta}\right)^{{n+1}\over{2}}
\int_0^\infty \frac{u^{{n-1}\over{2}} \left(\frac{u}{\zeta}+1\right)^m \rd \vartheta}{\re^{\alpha_*+u}-\varsigma} \nonumber\\
&\approx \frac{1}{2}\left(\frac{2}{\zeta}\right)^{{n+1}\over{2}} \bigg[
\int_0^\infty \frac{u^{{n-1}\over{2}} \rd \vartheta}{\re^{\alpha_*+u}-\varsigma}
+ \frac{m}{\zeta} \int_0^\infty \frac{u^{{n+1}\over{2}}
\rd \vartheta}{\re^{\alpha_*+u}-\varsigma}
+ \frac{m(m-1)}{2\zeta^2} \int_0^\infty \frac{u^{{n+3}\over{2}}
\rd \vartheta}{\re^{\alpha_*+u}-\varsigma} \bigg] \nonumber\\
&\approx \frac{1}{2}\left(\frac{2}{\zeta}\right)^{{n+1}\over{2}}
\varsigma\bigg[\Gamma\left(\frac{n+1}{2}\right)\Li_{\frac{n+1}{2}}\left(\varsigma\re^{-\alpha_*}\right)\nonumber\\
&~~~~+ \frac{m}{\zeta}\Gamma\left(\frac{n+3}{2}\right)\Li_{\frac{n+3}{2}}\left(\varsigma\re^{-\alpha_*}\right)
+\frac{m(m-1)}{2\zeta^2}\Gamma\left(\frac{n+5}{2}\right)
\Li_{\frac{n+5}{2}}\left(\varsigma\re^{-\alpha_*}\right)
\bigg],
\end{align}
therefore, $\mathcal J_{n,m} = -\dfrac{\partial}{\partial\alpha} J_{n,m}
= -\dfrac{\partial}{\partial\alpha_*} J_{n,m}$ becomes
\begin{align*}
\mathcal J_{n,m}\approx \, & \frac{1}{2}\left(\frac{2}{\zeta}\right)^{\frac{n+1}{2}}
\varsigma\bigg[\Gamma\left(\frac{n+1}{2}\right)\Li_{\frac{n-1}{2}}\left(\varsigma\re^{-\alpha_*}\right)\\
&+ \frac{m}{\zeta}\Gamma\left(\frac{n+3}{2}\right)\Li_{\frac{n+1}{2}}\left(\varsigma\re^{-\alpha_*}\right)
+ \frac{m(m-1)}{2\zeta^2}\Gamma\left(\frac{n+5}{2}\right)\Li_{\frac{n+3}{2}}\left(\varsigma\re^{-\alpha_*}\right)
\bigg].
\end{align*}
The Lorenz number is then simplified to be
\begin{align}
L \approx \frac{d+2}{2}
\left[\frac{d+4}{2}\frac{\Li_\frac{d+4}{2} \left(\varsigma\re^{-\alpha_*}\right)}
{\Li_\frac{d}{2}\left(\varsigma\re^{-\alpha_*}\right)}
-\frac{d+2}{2}\frac{\Li_\frac{d+2}{2}\left(\varsigma\re^{-\alpha_*}\right)^2}
{\Li_\frac{d}{2}\left(\varsigma\re^{-\alpha_*}\right)^2}\right].     \label{LLNR}
\end{align}
On the other hand, in the ultra-relativistic limit $\zeta \ll 1$,
\begin{align*}
J_{n,m} &\approx\left(\frac{1}{\zeta}\right)^{m+n}
\int_0^\infty \frac{u^{m+n-1} \rd \vartheta}{\re^{\alpha_*+u}-1}
=\left(\frac{1}{\zeta}\right)^{m+n}\Gamma(m+n)\,\Li_{m+n}(\re^{-\alpha_*}).
\end{align*}
In this case,
\begin{align*}
\mathcal J_{n,m} = -\frac{\partial}{\partial\alpha} J_{n,m}
\approx \left(\frac{1}{\zeta}\right)^{m+n}\Gamma(m+n)\,\Li_{m+n-1}(\re^{-\alpha_*}),
\end{align*}
and correspondingly,
\begin{align}
L \approx d\left[(d+1)\frac{\Li_{d+1} \left(\varsigma\re^{-\alpha_*}\right)}
{\Li_{d-1}\left(\varsigma\re^{-\alpha_*}\right)}
-d\frac{\Li_d\left(\varsigma\re^{-\alpha_*}\right)^2}
{\Li_{d-1}\left(\varsigma\re^{-\alpha_*}\right)^2}\right].   \label{LLUR}
\end{align}
Finally, eqs.\eqref{LLNR} and \eqref{LLUR} can be summarized
in the unified form \eqref{LL}.

\section{Universal $L$ for strongly degenerate Fermi gas}

Here we outline the procedure for obtaining the universal result for the Lorenz number
of strongly degenerate Fermi gas.
We first introduce the variables $u = \zeta(\cosh\vartheta -1)$ and $u_0 = - \alpha_*$.
Then we can write
\begin{align*}
J_{n,m} &= \int_0^\infty
\frac{\sinh^n \vartheta \cosh^m \vartheta \rd \vartheta}
{\re^{\alpha + \zeta\cosh\vartheta}+ 1}
= \int_0^\infty \frac{\rd\varphi(u)}{\re^{u - u_0}+1},
\end{align*}
where
\begin{align*}
\varphi(u) = \frac{1}{n+1} \sinh^{n+1}\left[\arcosh\left(\frac{u+\zeta}{\zeta}\right)\right]
\mathcal F_{\frac{1-m}{2},\frac{n+1}{2}}\left[
\arcosh\left(\frac{u+\zeta}{\zeta}\right)\right],
\end{align*}
and $\mathcal F_{a,b}(\vartheta)$ is the following Gaussian hypergeometric function
\begin{align*}
\mathcal F_{a,b}(\vartheta) \equiv
\,_2F_1\left(a,b;b+1; -\sinh^2 \vartheta\right).
\end{align*}

According to Sommerfeld's lemma, the following integral
can be approximately expressed as
\begin{align*}
\int_0^\infty \frac{\rd \varphi(u)}{\re^{u-u_0}+1}
\approx \varphi(u_0) + \frac{\pi^2}{6}
\varphi''(u_0)
+ \cdots,
\end{align*}
provided $\varphi(0) = 0$, $\displaystyle\lim_{u\to\infty}\frac{\varphi(u)}
{\re^{u-u_0}+1} = 0$ and $u_0 \gg 1$. Therefore, we have
\begin{align*}
J_{n,m} \approx \varphi(-\alpha_*) + \frac{\pi^2}{6}\varphi''(-\alpha_*),
\end{align*}
and further,
\begin{align*}
\mathcal J_{n,m} = -\frac{\partial}{\partial\alpha} J_{n,m}
\approx \varphi'(-\alpha_*) + \frac{\pi^2}{6}\varphi'''(-\alpha_*),
\end{align*}
where
\begin{align*}
\varphi'(-\alpha_*)
&= \frac{1}{\zeta}\left(-\frac{\alpha}{\zeta}\right)^m
\sinh^{n-1}\left[\arcosh\left(-\frac{\alpha}{\zeta}\right)\right],\\
\varphi'''(-\alpha_*)
&= \frac{\pi^2}{6}\frac{1}{\zeta^3}
\bigg\{(m+n-1)(m+n-2)\left(-\frac{\alpha}{\zeta}\right)^{m+2}
+ m(m-1)\left(-\frac{\alpha}{\zeta}\right)^{m-2}\\
&~~~~~~~~~~~~~ -\left[2 m (m+n-2)+n-1\right]
\left(-\frac{\alpha}{\zeta}\right)^m \bigg\}
\sinh^{n-5}\left[\arcosh\left(-\frac{\alpha}{\zeta}\right)\right].
\end{align*}
Inserting the above approximate result for $\mathcal J_{n,m}$ into
eq.\eqref{Lorenz} and making some further simplifications,
we get the desired result \eqref{LFermi}.

\section*{Acknowledgement}

This work is supported by the National Natural Science Foundation of China under the grant
No. 12275138 and by the Hebei NSF under the Grant No. A2021205037.

\section*{Data Availability Statement}

This work is purely theoretical and contains only analytic analysis.
Hence there is no associated numeric data.

\section*{Declaration of competing interest}

The authors declare no competing interest.


\providecommand{\href}[2]{#2}\begingroup
\footnotesize\itemsep=0pt
\providecommand{\eprint}[2][]{\href{http://arxiv.org/abs/#2}{arXiv:#2}}

\endgroup

\end{document}